\documentclass[12pt]{article}
\usepackage{amsmath,euscript}
\usepackage{graphicx}
\usepackage{bm}
\usepackage{epsfig}

 \newcount\hour \newcount\minute
  \hour=\time \divide \hour by 60
  \minute=\time
  \newcommand{\mydate}{\ \today \ - \number\hour :\ifnum \minute<10 0\fi 
\number\minute}

\begin{document}
\newcommand{\ltap}{\stackrel{<}{_\sim}}
\newcommand{\gtap}{\stackrel{>}{_\sim}}
\newcommand{\gsim}{{~\raise.15em\hbox{$>$}\kern-.85em
          \lower.35em\hbox{$\sim$}~}}
\newcommand{\lsim}{{~\raise.15em\hbox{$<$}\kern-.85em
          \lower.35em\hbox{$\sim$}~}}
\begin{titlepage}
\begin{flushright}
{\tt hep-ph/0407076}\\[0.2cm]
July 6, 2004
\end{flushright}

\vspace{0.7cm}
\begin{center}
\Large\bf 
Right-handed currents, $C\!P$ violation, and $B \to VV$
\end{center}

\vspace{0.8cm}
\begin{center}
{\sc Alexander L. Kagan}\\
\vspace{0.7cm}
{\sl Department of Physics\\
University of Cincinnati\\
Cincinnati, Ohio 45221, U.S.A.}
\end{center}

\vspace{1.0cm}
\begin{abstract}
\vspace{0.2cm}\noindent
Precision CP violation measurements in rare hadronic 
$B$ decays could
provide clean signatures of parity symmetric new physics, 
implying the existence of
$SU(2)_L \times  SU(2)_R \times U(1)_{B-L} \times  P$ symmetry at 
high energies.
New contributions to the weak scale Hamiltonian which respect parity to $O(1\%)$ in supersymmetric realizations are compatible with an $SU(2)_R$ breaking scale $M_R \le M_{\rm GUT}$, implying that sensitivity to the GUT scale is possible.  
The generic case of right-handed currents without left-right symmetry is also discussed. 
A detailed analysis of $B \to VV$ polarization in QCD factorization reveals that 
the low longitudinal polarization fraction $f_L (\phi K^*) \approx 50\%$ can be accounted for in the SM via a QCD penguin annihilation graph.
The ratio of transverse rates
$\Gamma_\perp / \Gamma_\parallel $ provides
a sensitive test for new right-handed currents. 
CP violation measurements in $B \to VV$ decays can discriminate between new contributions to the dipole and four quark operators.  
\end{abstract}
\vspace{.75cm}
\center{Contribution to {\it Discovery Potential of a High Luminoscity Asymmetric $B$ Factory}, eds. J. Hewett and D. Hitlin}
\vfil

\end{titlepage}

\section{Introduction}
\label{sec:first}

In this contribution we discuss signals for right-handed currents in rare hadronic $B$ decays.  Signals in radiative $B$ decays are reviewed elsewhere in this report.
Implications of right-handed currents  for $C\!P$-violation phenomenology are addressed in 
$SU(2)_L \times SU(2)_R \times U(1)_{B-L} \times P$
symmetric models, and in the more general case of no left-right symmetry.
We will see that it may be possible to distinguish between these scenarios  
at a high luminosity $B$ factory. 
Remarkably, the existence of 
$SU(2)_R $ symmetry could be inferred even if  it is broken at a scale many orders of magnitude 
larger than the weak scale, e.g., $M_R \lsim M_{\rm GUT} $, in parity symmetric models 
\cite{superBtalk,engelkagan}.
An explicit supersymmetric realization is briefly described.

A direct test for right-handed currents from polarization measurements in $B$ decays to light vector meson pairs is also discussed~\cite{kaganVV}.
In the event that non-Standard Model $C\!P$-violation is confirmed, 
e.g., in the $B \to \phi K_s $ time-dependent $C\!P$ asymmetry, an important
question will be whether it arises via New Physics contributions to the four-quark operators, the $b \to sg$ dipole operators, or both.
We will see that this question can be addressed by comparing $C\!P$ asymmetries
in the different transversity final states in pure penguin $B \to VV$ decays, e.g., $B \to \phi K^*$.  The underlying reason is large suppression of 
the {\it transverse} dipole operator matrix elements.
It is well known that it is difficult to obtain new
${\cal O}(1)$ $C\!P$ violation effects at the {\it loop-level} from the 
{\it dimension-six} four-quark operators.  Thus, this information
could help discriminate between scenarios in which New Physics effects are induced via loops versus at tree-level.

\begin{figure}[htb]
\centerline{
\hbox{\hspace{0.2cm}
\includegraphics[width=9truecm]{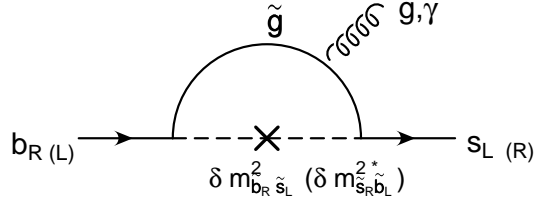}}}
{\caption[1]{\label{fig:susyloops} Down squark-gluino loop contributions to
the Standard Model and opposite chirality dipole operators in the squark mass insertion approximation.}}
\end{figure}

Extensions of the Standard Model often include new $b \to s_R$ right-handed currents.   
These are conventionally associated with opposite chirality effective operators $\tilde{Q}_i$ which are
related to the Standard Model operators $Q_i$ by parity transformations,
\vspace{-.2cm}
\begin{itemize}
\item QCD Penguin operators
\vspace{-.3cm}    
\[\begin{array}{ll}
\hspace{-.5cm} Q_{3,5} = (\bar s  b)_{V-A}\, (\bar q q )_{V\mp A} &\rightarrow \,
\tilde{Q}_{3,5} =(\bar s  b)_{V+A}\, (\bar q q )_{V\pm A}  \\
\hspace{-.5cm} Q_{4,6} = (\bar s_i  b_j)_{V-A}\, (\bar q_j q_i )_{V\mp A} &\rightarrow \,
\tilde{Q}_{4,6} =(\bar s_i  b_j)_{V+A}\, (\bar q_j q_i )_{V\pm A} 
\end{array}\]

\item Chromo/Electromagnetic Dipole Operators
\[\begin{array}{ll}
\hspace{-.5cm}Q_{7\gamma} = \frac{e}{8\pi^2} m_b 
\bar s_i \sigma^{\mu\nu} (1 + \gamma_5 ) b_i F_{\mu\nu} &\rightarrow \,
\tilde{Q}_{7\gamma} = \frac{e}{8\pi^2} m_b 
\bar s_i \sigma^{\mu\nu} (1 - \gamma_5 ) b_i F_{\mu\nu}\\
\hspace{-.5cm}Q_{8g} = \frac{g_s}{8\pi^2} m_b
\bar s \sigma^{\mu\nu} (1 + \gamma_5 ) t^a b G^a_{\mu\nu} &\rightarrow \,
\tilde{Q}_{8g} =  \frac{g_s}{8\pi^2} m_b \bar s \sigma^{\mu\nu} (1 - \gamma_5 ) t^a b G^a_{\mu\nu} 
\end{array}\]

\item Electroweak Penguin Operators
\[\begin{array}{ll}
\hspace{-.5cm} Q_{7,9} = \frac{3}{2} ( \bar s  b)_{\rm V-A}
\,  e_{q}\, ( \bar q q )_{\rm V\pm A} &\rightarrow\, \tilde{Q}_{7,9} = \frac{3}{2} ( \bar s b)_{\rm V+A}   \,e_{q} \,( \bar q q )_{\rm V\mp A} \\
\hspace{-.5cm} Q_{8,10} = \frac{3}{2} ( \bar s_{i} b_{j} )_{\rm V-A}\,
       e_{q} \,( \bar q_{j}  q_{i})_{\rm V\pm A} & \rightarrow \,
\tilde{Q}_{8,10} = \frac{3}{2} ( \bar s_{i} b_{j} )_{\rm V+A}\,
 \,e_{q} \,( \bar q_{j}  q_{i})_{\rm V\mp A}   \end{array}\]
\end{itemize}
Examples of New Physics which could give rise to right-handed currents
include supersymmetric loops which contribute to the 
QCD penguin or chromomagnetic dipole operators.  These are discussed
in detail elsewhere in this report.  
Figure \ref{fig:susyloops} illustrates the well known squark-gluino loops
in the squark mass-insertion approximation.
For example, the down-squark mass-insertion $\delta m^2_{\tilde{b}_R \tilde{s}_L }$ ($\delta m^{2\,*}_{\tilde{s}_R \tilde{b}_L }$) would contribute to $Q_{8g}$ ($\tilde{Q}_{8g} $), whereas
$\delta m^2_{\tilde{b}_L \tilde{s}_L }$ ($\delta m^{2}_{\tilde{s}_R \tilde{b}_R }$)
would contribute to
$Q_{3,..6}$ ($\tilde{Q}_{3,..,6}$).
Right-handed currents could also arise at tree-level via new contributions to 
the QCD or electroweak penguin operators, e.g., due to flavor-changing
$Z^{(\prime)}$ couplings, $R$-parity violating couplings, or color-octet exchange.  

\section{Null Standard Model $C\!P$ asymmetries}

We exploit the large collection of {\it pure-penguin} $B \to f$ decay modes, which in the Standard Model have
\begin{itemize}
\vspace{-.2cm}
\item null decay rate $C\!P$-asymmetries, $A_{C\!P} (f) \sim 1\%$, or
\item null deviations of the time-dependent $C\!P$-asymmetry coefficient
$S_{f_{\rm CP}} $ from $(\sin 2\beta)_{J/\Psi K_s} $ in decays to $C\!P$-eigenstates, $|(\sin 2\beta)_{J/\Psi K_s } + (-)^{\rm CP} {S_{f_{\rm CP}}}| \sim 1\%$, or 
\item null triple-product $C\!P$-asymmetries $A_T^{0,\parallel} (f)\sim 1\%$ in $B \to VV$ decays.
\end{itemize}

We recall that there are three helicity amplitudes 
${\cal \bar A}^{ h}$ ($h=0,-,+$) in $\bar B\to VV$ decays:
${\cal \bar A}^{ 0}$, in which both vectors are longitudinaly polarized;
${\cal \bar A}^{ -}$, in which both vectors have negative helicity; and 
${\cal \bar A}^{+}$, in which both vectors have positive helicity.
In the transversity basis~\cite{dunietz}, the amplitudes are given by,
\begin{equation}\label{transversity} {\cal \bar A}_{\perp,\parallel} = ({\cal \bar A}^- \mp {\cal \bar A}^+ )/\sqrt{2},~~~{\cal \bar A}_{0}= {\cal\bar A}^0
\end{equation}
In $B$ decays, ${\cal A}_{\perp,\parallel} = ({\cal  A}^+ \mp {\cal  A}^-)/\sqrt{2}$.
The CP-violating triple-products~\cite{Valencia} (related to $\vec q \cdot  \vec{\epsilon}_1 \times \vec{\epsilon}_2 $) are then given by
\begin{equation}
A_T^{0 \,(\parallel) }= \frac12 \left( \frac{ {\rm Im}({\cal \bar A}_{\perp \, (\parallel)}{\cal \bar A}_0^* )}{
\sum |{\cal \bar A}_{i} |^2 }  - \frac{ {\rm Im}({\cal  A}_{\perp\, (\parallel)} {\cal  A}_0^* )}{
\sum |{\cal A}_{i} |^2 } \right) \,.
\label{AT}\end{equation}
The triple-products are discussed in detail in the contribution of
A.~Datta.

A partial list of null Standard Model $C\!P$ asymmetries in pure-penguin decays
is given below \cite{grossmanworah},
\begin{itemize}
\vspace{-.2cm}
\item   $A_{CP} (K^0 \pi^\pm )$, $A_{CP} (\eta^\prime K^\pm )$, 
$A_{CP}(\phi K^{*0,\pm})_{0,\parallel,\perp}$, $A_{CP} (K^{*0} \pi^\pm )$, 
$A_{CP} (K^{*0} \rho^\pm )_{ 0,\parallel,\perp}$, $A_{CP} (K_1 \pi^\pm )$, $A_{CP} (K^0 a_1^\pm )$, $A_{CP} (\phi K^{0, \pm} )$,...
\item $S_{\phi K_s }$, $S_{\eta^\prime K_s }$, $(S_{\phi K^{*0}})_{ 0,\parallel,\perp}$,
$(S_{\phi K_1 })_{0,\parallel,\perp}$, $S_{K_s K_s K_s }$,...
\item $A_T^{0,\parallel} (\phi K^{* 0,\pm})$, $A_T^{0,\parallel} (K^{* 0} \rho^\pm )$,...
\end{itemize}
In addition, there are several modes which are penguin-dominated and are predicted to have approximately null or small Standard Model asymmetries, e.g., $S_{K^+ K^- K^0 }$ ($\phi$ subtracted)~\cite{BelleKKK,grossmanetal},
$S_{K_s \pi^0 }$~\cite{GGR}, and  $S_{f^0 K_s }$.

\section{Right-handed currents and $C\!P$-violation}

Under parity, the effective operators transform as $Q_{i} \leftrightarrow \tilde{Q}_{i}$. The New Physics amplitudes,
for final states $f$ with parity $P_f$, therefore satisfy 
\begin{equation}\label{ANPparity}
\hspace{-1cm} \langle f | Q_i |B\rangle  = -(-)^{P_f}  \langle f | \tilde{Q}_i |B\rangle \,\Rightarrow\, A^{NP}_i (B\rightarrow f) \propto C_{i}^{\rm NP} (\mu_b) -(-)^{P_f} \tilde{C}_i^{\rm NP} (\mu_b)\,,
\end{equation}
where $C_i^{\rm NP} $ and $\tilde{C}_i^{\rm NP}$ are the new Wilson
coefficient contributions to the i'th pair of Standard Model and opposite chirality operators~\cite{SSI}.
It follows that for decays to $PP$, $VP$, and $SP$ final states,
where $S$, $P$ and $V$ are scalar, pseudoscalar, and vector mesons,
respectively, the New Physics amplitudes satisfy
\[
A^{\rm NP}_i (B\rightarrow PP)  \propto C_{i}^{\rm NP} (\mu_b) -\tilde{C}_i^{\rm NP} (\mu_b),~~~
A^{\rm NP}_i (B\rightarrow VP) \propto C_{i}^{\rm NP} (\mu_b) +\tilde{C}_i^{\rm NP} (\mu_b)\]
\begin{equation}
A^{\rm NP}_i (B\rightarrow SP) \propto C_{i}^{\rm NP} (\mu_b) +\tilde{C}_i^{\rm NP} (\mu_b)\label{ANPPP}\,.
\end{equation}
In $B \to VV$ decays the $\perp$ transversity and $0,\,\parallel$
transversity final states are $P$-odd and $P$-even, respectively, yielding
\begin{equation}
 A^{NP}_i (B\rightarrow VV )_{ 0,\parallel} \propto C_{i}^{\rm NP} (\mu_b) -\tilde{C}_i^{\rm NP}
(\mu_b) ,~~~
A^{NP}_i (B\rightarrow VV)_{ \perp} \propto C_{i}^{\rm NP} (\mu_b) +\tilde{C}_i^{\rm NP} (\mu_b)\label{ANPVV}\,.
\end{equation}
Similarly, replacing one of the vector mesons with an axial-vector meson
gives
\begin{equation}
A^{NP}_i (B\rightarrow VA )_{ 0,\parallel} \propto C_{i}^{\rm NP} (\mu_b) +\tilde{C}_i^{\rm NP} (\mu_b),~~~
 A^{NP}_i (B\rightarrow VA)_{ \perp}  \propto C_{i}^{\rm NP} (\mu_b) -\tilde{C}_i^{\rm NP} (\mu_b)\,. \label{ANPAV}
 \end{equation}   
It is useful to classify the null and approximately null Standard Model
$C\!P$ asymmetries listed above according to whether the final state
is $P$-odd or $P$-even,
\begin{itemize}
\vspace{-.2cm}
\item P-even: $A_{CP} (K^0 \pi^\pm )$, $A_{CP} (\eta^\prime K^\pm )$, 
$A_{CP}(\phi K^{*\pm})_{ 0,\parallel}$, $S_{\eta^\prime K_s }$,
$(S_{\phi K^{*0}})_{ 0,\parallel}$,
$A_{CP} (K^{*0} \rho^\pm )_{ 0,\parallel}$, 
\newline $A_{CP} (K_1 \pi^\pm )$, 
$A_{CP} (K^0 a_1^\pm )$,
$(S_{\phi K_1 })_{ \perp}$,... 
\item P-odd: $A_{CP} (\phi K^{\pm} )$, $S_{\phi K_s }$, $A_{CP} (K^{*0} \pi^\pm )$, 
$A_{CP} (\phi K^{*\pm})_{ \perp}$,
$(S_{\phi K^{*0} })_{ \perp}$,
$(S_{\phi K_1 } )_{ 0,\parallel}$,...
\item Modes with small Standard Model asymmetries:
$S_{K^+ K^- K^0 }$ (approximately $P$-even),
$S_{K_s \pi^0 }$ (P-even), and $S_{f^0 K_s }$ (P-odd).
\end{itemize}
We are now ready to discuss implications for $C\!P$ violation
phenomenology in the two classes of models mentioned earlier.

\subsection{Parity symmetric New Physics}
In the limit in which New Physics is parity symmetric at the weak scale
the relation $C_i^{\rm NP} (\mu_W ) = \tilde{C}_i^{\rm NP} (\mu_W) $
would hold.
In light of  (\ref{ANPparity}) this would imply \cite{SSI,superBtalk}
\begin{itemize}
\vspace{-.2cm}
\item preservation    
of null $C\!P$ asymmetry predictions in {\it P-even} final states.  Similarly,
the $\epsilon^\prime /\epsilon $ constraint would be trivially satisfied.
\item possibly {large departures} from null $C\!P$ asymmetries in {\it  P-odd} final states.
\end{itemize}
\vspace{-.1cm}
For example, no deviations in 
$S_{\eta^\prime K_s}$, $(S_{\phi K^{*0}})_{0, \parallel} $, $A_{CP} (\phi K^{*\pm})$, $A_{CP} (K^0 \pi^\pm)$ could be accompanied by significant
deviations in $S_{\phi K_s}$, $A_{CP} (\phi K^\pm)$,
$(S_{\phi K^{*0}})_{\perp} $,  and $S_{f^0 K_s} $.  
Both of the triple-products $A_T^{0}$ and $A_T^{\parallel}$ in (\ref{AT})
could be affected through a modification of  ${\cal A_{\perp}}(VV)$.
However, there would be no novel $C\!P$ asymmetry in the interference of 
the parallel and longitudinal polarizations.  Equivalently, 
the measurable quantities $\Delta_{0}$ and $\Delta_{\parallel}$ defined below
\begin{equation}
\Delta_{0\,(\parallel)} =( {\rm Arg }\,{\cal \bar A}_{0\, (\parallel )}  - {\rm Arg }\,{\cal \bar A}_{\perp} )- ( {\rm Arg }\,{\cal  A}_{0\, (\parallel )}  - {\rm Arg }\,{\cal A}_{\perp} ) 
\end{equation}
would be equal.

Parity-symmetric new physics requires $SU(2)_L \times SU(2)_R \times U(1)_{B-L}\times P$ symmetry at high energies.
Thus, exact weak scale parity can not be realized due to
renormalization group effects below the
$SU(2)_R$ breaking scale,
 $M_R$.  
 Potentially, the largest source of parity violation is the difference between the top and bottom quark Yukawa couplings. 
In particular,  when $\lambda^t \ne \lambda^b$ the charged Higgs Yukawa couplings break parity.  Two scenarios for the Yukawa couplings naturally present themselves:
\begin{itemize}
\item moderate $\tan\beta$, or $\lambda^t >>\lambda^b$
\item {maximal-parity}:
 {$\lambda^b = \lambda^t + {\cal O}(V_{cb} )$ or $\tan\beta  \cong m_t /m_b $}
Small corrections to the limit of equal up and down Yukawa matrices 
are required in order to generate the observed CKM quark mixings and light quark masses.
$V_{cb}$ therefore sets the scale for minimal parity violation in the Yukawa sector.
\end{itemize}

A large hierarchy between the $SU(2)_R$ breaking scale and the weak scale 
can be realized naturally in supersymmetric left-right symmetric models.  
These models contain
two Higgs bidoublet superfields $\Phi_{1,2} (2_L , 2_R , 0_{B-L} )$ (or four $SU(2)_L$
doublets).
Via the `doublet-doublet splitting' mechanism \cite{BDM} two linear combinations of the Higgs doublets acquire masses of order $M_R$, leaving the two light Higgs doublets of the 
MSSM. 
Realization of approximately parity symmetric contributions to the 
{\it dipole} operators favors explicit $C\!P$ violation.
Spontaneous $C\!P$ violation
could lead to complex $P$-violating vacuum expectation values
which would feed into new loop contributions to the operators.  For example, $P$ invariance above the 
weak scale would imply
\begin{equation}
C_{8g}^{\rm NP}  = \kappa \langle {\phi}\rangle,~~~~~~~~
\tilde{C}_{8g}^{\rm NP} = \kappa \langle{\phi^\dagger}\rangle,
\label{dipoleparity}\end{equation}
where $\langle{\phi}\rangle $ breaks $SU(2)_L$ and $\kappa \sim 1/M_{\rm NP}^2 $ 
is in general complex due to explicit $C\!P$ violating phases.  
($M_{\rm NP}$ is an order TeV new physics scale, e.g., the squark or gluino masses in Figure 1).  Thus,  $\langle{\phi}\rangle$ would have to be real to good approximation in order to obtain $C_{8g}^{\rm NP} \approx \tilde{C}_{8g}^{\rm NP} $.
Note that this also requires real gaugino masses; otherwise RGE effects would
induce a complex Higgs bilinear $B$ term in the scalar potential, thus leading to complex $\langle{\phi}\rangle $.  Ordinary parity symmetry
insures real $U(1)_{B-L}$ and $SU(3)_C$ gaugino masses.  
Real $SU(2)_L \times SU(2)_R $ gaugino masses
naturally follow from the $SO(10)$ generalization of parity
\cite{mohapatra}.  All the VEVs entering new {\it four-quark} operator loops can, in principle, be parity neutral.
Therefore, real VEVs are less crucial for obtaining approximately parity-symmetric four-quark operator 
contributions.

We have carried out a two-loop RGE analysis for down squark-gluino loop contributions 
to the dipole operators.  Choosing parity symmetric boundary conditions at 
$M_R$, taking $M_R \le M_{\rm GUT}$, and running to the weak scale we obtain 
\begin{itemize}
\item {Moderate $\tan\beta$}, e.g. , $\tan \beta \sim 5$:
\[{\tiny  \frac{{\rm Re}[C_{8g}^{\rm NP}(m_{W} ) - \tilde{C}^{\rm NP}_{8g}(m_{W} ) ]}
{{\rm Re }[C^{\rm NP}_{8g}(m_{W} ) +\tilde{C}^{\rm NP}_{8g}(m_{W} ) ]} \le 10\% },\qquad
{\tiny  \frac{{\rm Im}[C_{8g}^{\rm NP}(m_{W} ) - \tilde{C}^{\rm NP}_{8g}(m_{W} ) ]}
{{\rm Im }[C^{\rm NP}_{8g}(m_{W} ) +\tilde{C}^{\rm NP}_{8g}(m_{W} ) ]} \le 10\% }\]
\item{Maximal parity}, $\tan\beta \cong m_t /m_b $
\[{\tiny  \frac{{\rm Im}[C_{8g}^{\rm NP}(m_{W} ) - \tilde{C}^{\rm NP}_{8g}(m_{W} ) ]}
{{\rm Im }[C^{\rm NP}_{8g}(m_{W} ) +\tilde{C}^{\rm NP}_{8g}(m_{W} ) ]} =O( 1\%) }\]
\end{itemize}
The above quantities give a measure of parity violation in the weak scale 
Wilson coefficients.  Thus, we see that
for $M_R \le  M_{\rm GUT}$, new $C\!P$ violating contributions to the
low energy Lagrangian could respect parity to ${\cal O}(1\% )$.
Precision $C\!P$ violation measurements in $B$ decays which respect {(violate)} null SM predictions in $P$-even ($P$-odd) final states would therefore provide {evidence for}
$SU(2)_L \times SU(2)_R \times U(1)_{B-L} \times P$ symmetry, even if 
$SU(2)_R$ is broken at the GUT scale.
Similar results are obtained for survival of parity in the four-quark operators
\cite{engelkagan}.

\subsubsection{The $^{199}$Hg mercury edm constraint}

Any discussion of dipole operator phenomenology must consider
the upper bound on the strange quark chromo-electric dipole moment $d_s^C$, obtained from the upper bound on the $^{199}$Hg mercury edm \cite{dscexp}.
Correlations between $d_s^C$ and new $C\!P$ violating contributions
to $C_{8g}$, $\tilde{C}_{8g}$ are most easily seen by writing the dipole operator
effective Hamiltonian in the weak interaction basis,
\begin{equation}\label{weakbasis} \frac{G_F}{\sqrt{2}} \,V_{cb} V_{cs} \,
{C_{i_L j_R} }\,\frac{g_s}{8\pi^2} m_b \, \bar i \, \sigma^{\mu \nu} (1+\gamma^5 )\, j \,G_{\mu\nu} +h.c.\,.
\end{equation}
$|{i_L }\rangle$ and $|{i_R}\rangle$ ($i=1,2,3$) are the left-handed and right-handed
down quark weak interaction eigenstates, respectively.
The mass eigenstates can be written as 
$ |{d^i_{L (R)} }\rangle = x^{L (R)}_{i j} |{i_{L (R)}}\rangle$,
where $d^{1,2,3} $ stands for the $d,s,b$ quarks, respectively,
and $x_{ii}^{L,R} \approx 1$.
The bound on $d_s^C$ is 
 ${\rm Im}\,C_{s_L s_R} \lsim  4 \times 10^{-4}$, with large theoretical uncertainty, where
$C_{s_L s_R}$ is the flavor diagonal strange quark dipole operator coefficient (in the mass eigentate basis).  
It is given as 
\begin{equation}\label{Css}
C_{s_L s_R} \approx  C_{2_L 2_R} + x_{23}^{L\,*} C_{3_L 2_R}+ x_{23}^R C_{2_L 3_R}
+ x_{23}^{L\,*} x_{23}^R C_{3_L 3_R} +...\,.
\end{equation}
Similarly, the $b \to sg$ Wilson coefficients are given as
\begin{equation}\label{C8gCij}
C_{8g} \approx  C_{2_L 3_R} + x_{23}^{L\,*}   C_{3_L 3_R}+...,~~~
\tilde{C}_{8g} \approx  C^*_{3_L 2_R} + x_{23}^{R\,*}   C^*_{3_L 3_R}+...
\end{equation}
If significant contributions to the CKM matrix elements
are generated in the down quark sector, then $x_{23}^L , x_{32}^L  \sim V_{cb} $,
$x^L_{13}, x^L_{31} \sim V_{ub}$, and $x^L_{12}, x^L_{21}\sim \theta_c $.
In the absence of special flavor symmetries, similar magnitudes
would be expected for the corresponding right-handed quark mixing coefficients, $x_{ij}^R$.  
Generically, we therefore expect $C_{s_L s_R} \sim V_{cb}\, C_{8g}$.
$S_{\phi K_s } <0$ would correspond to ${\rm Im}\, [C_{8g} (m_b )+ \tilde{C}_{8g} (m_b ) ]\sim 1 $.  Thus, ${\cal O}(1)$
$C\!P$ violating effects generically correspond to a value for $d_s^C$ which is a factor of 100 too large.
One way to evade this bound is by invoking some mechanism, e.g., flavor symmetries, for generating the large hierarchies
$x_{23}^R <<  x_{23}^L$ and $C_{3_L 2_R} << C_{2_L 3_R}$.
An elegant alternative is provided by parity symmetry \cite{win03}.
It is well known that edm's must vanish in the parity symmetric limit, see e.g. \cite{mohapatra}.
For example, in (\ref{Css}) exact parity would imply $x_{23}^{L}  =
x_{23}^R $, $C_{3_L 2_R} = C_{2_L 3_R}^* $ and real $C_{i_L i_R}$, thus yielding a real coefficient, $C_{s_L s_R}$.
An RGE analysis along the lines discussed above is required in order to determine the extent to which this can be realized at low energies.  
We find that in both the maximal parity scenario ($\tan\beta \cong m_t /m_b $)
and in moderate $\tan\beta$ scenarios it is possible to obtain $S_{\phi K_s } <0$ and at the same time satisfy the   
bound on $d_s^C$ if $M_R \le M_{\rm  GUT}$ \cite{engelkagan}.

\subsection{Generic case:  Right-handed currents without Parity}

In the parity-symmetric scenario, an unambiguous theoretical interpretation of 
the pattern of $C\!P$ violation is possible because null predictions
are maintained for the $P$-even final states. 
However, if new contributions to the $Q_i$ and $\tilde{Q}_i$ operators are unrelated,
then $C\!P$ asymmetries in the $P$-odd and $P$-even null Standard Model modes could differ significantly both {\it from each other, and from the null predictions}.   This is due to the opposite relative sign between the left-handed and right-handed New Physics amplitudes for $P$-odd and $P$-even final states in Eqs. (\ref{ANPparity})--
(\ref{ANPAV}).
For example, $S_{\phi K_s }$ and $S_{\eta^\prime K_s }$ could be affected differently 
in the MSSM \cite{khalil,hou}.
An interesting illustration would be provided by models with 
$O(1)$ contributions to the $\tilde{Q}_i $,
and negligible new contributions to the $Q_i$.  
This could happen, for example, in supersymmetric models 
with large (negligible) $\tilde{s}_{R (L)} -\tilde{b}_{R (L)} $ squark mixing
\cite{hou}--\cite{Chang}, or in models in which $R$-parity violation 
induces opposite chirality four-quark operators at the tree-level \cite{Datta}.
Unrelated right-handed currents could also arise
in warped extra dimension models with bulk left-right symmetry
\cite{burdman}.

\begin{figure}[htb]
\centerline{
\hbox{$\begin{array}{cc}
\includegraphics[width=7.1cm]{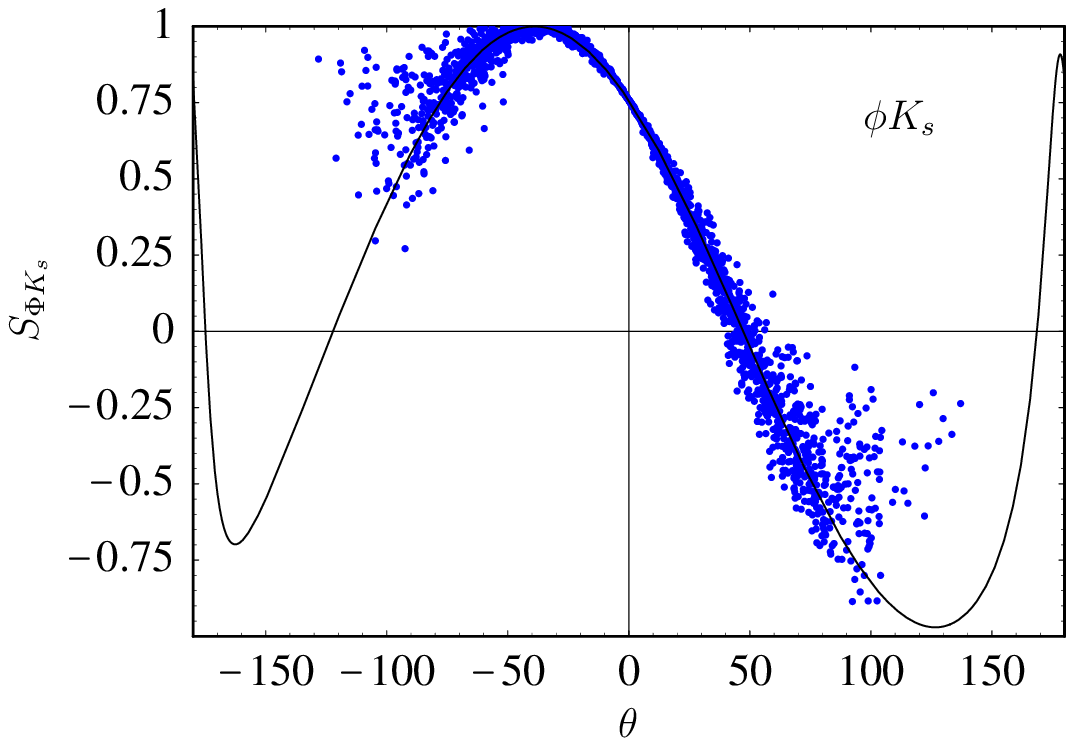} &
\hspace{.5cm}\includegraphics[width=7.1cm]{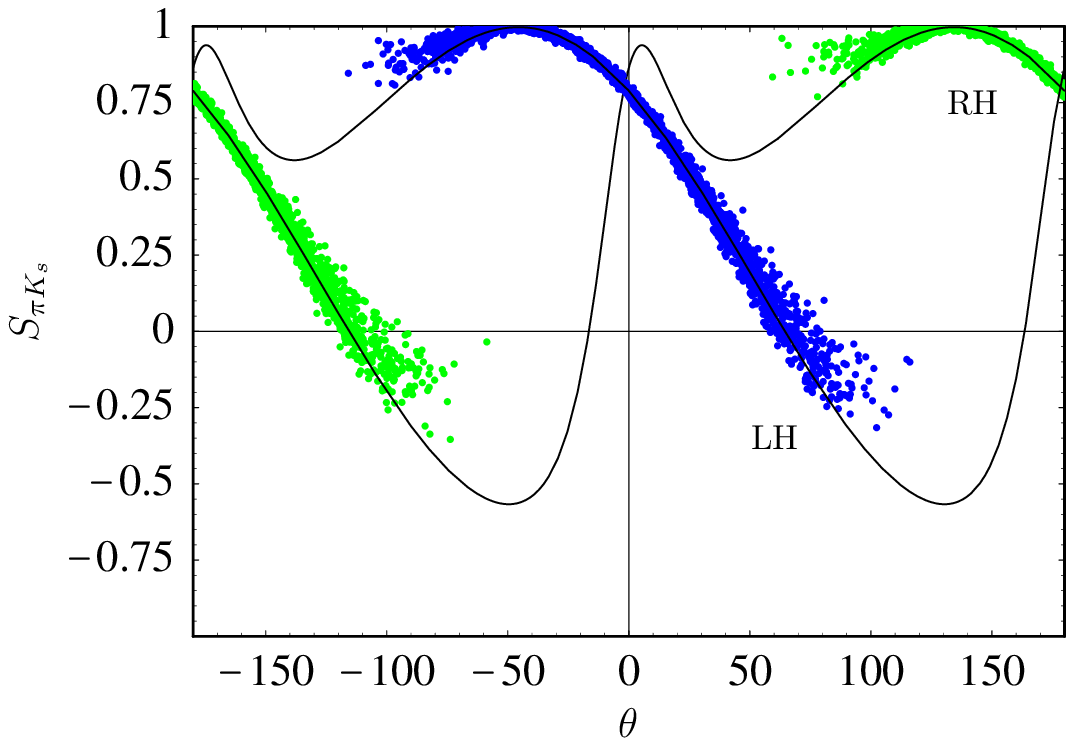}
\end{array}$}}
{\caption[1]{\label{fig:CPcomparison} Scatter plots in QCD factorization for $S_{\phi K_s}$ vs. $\theta$ for $C^{\rm NP}_{8g} (m_b ) +
\tilde{C}^{\rm NP}_{8g} (m_b) = e^{i\theta }$, and for 
$S_{\pi^0 K_s }$ versus $\theta$ for left-handed currents, $C^{\rm NP}_{8g} (m_b ) = e^{i\theta },\,\tilde{C}^{\rm NP}_{8g} =0$ (blue), and for right-handed-currents, $C^{\rm NP}_{8g} =0,\,\tilde{C}^{\rm NP}_{8g} (m_b ) = e^{i\theta }$ (green).}}
\end{figure}

Unfortunately, $C\!P$ asymmetry predictions 
have large theoretical uncertainties due to $1/m$ power corrections, especially 
from the QCD penguin annihilation amplitudes.  They are therefore difficult to interpret.
An illustration is provided in Figure \ref{fig:CPcomparison},
which compares predictions for  $S_{\phi K_s} $ and $S_{\pi^0 K_s }$
arising from new contributions to $Q_{8g}$ and $\tilde{Q}_{8g}$ in QCD factorization
\cite{BBNS3,benekeneubert}.
For $S_{\phi K_s}$ we take  
$C^{\rm NP}_{8g}(m_W)  + \tilde{C}^{\rm NP}_{8g} (m_W) = e^{i \theta} $.
For $S_{\pi^0 K_s} $ two corresponding cases are considered: 
(a) a purely left-handed current,  $C^{\rm NP}_{8g}(m_W)  = e^{i \theta} $, $ \tilde{C}^{\rm NP}_{8g} (m_W) = 0 $, (b) a purely right-handed current, ${C}^{\rm NP}_{8g} (m_W) = 0 $, $\tilde{C}^{\rm NP}_{8g}(m_W)  = e^{i \theta} $.
The scatter plots scan over the input parameter ranges given in 
\cite{benekeneubert} (with the exception of the Gegenbauer moments of the light meson light-cone distribution amplitudes and 
$m_c /m_b$, which have been set to their default values).
In addition, the branching ratios are required to lie within their 90\%
c.l. intervals.

Clearly, very different values for 
the two $C\!P$ asymmetries can be realized if the New Physics only appears in  $Q_{8g}$.  For example, for $\theta \sim 50^\circ$, it is possible to obtain $S_{\phi K_s } \sim -0.35$ 
and $S_{\pi^0 K_s} \sim 0.4 $.   The theoretical uncertainty
in $S_{\eta^\prime K_s}$
is larger than for $S_{\pi^0 K_s}$.  We therefore expect that even larger 
differences are possible between  $S_{\eta^\prime K_s}$ and $S_{\phi K_s }$,
for purely left-handed currents.
However, Figure  \ref{fig:CPcomparison} suggests that $S_{\phi K_s } < 0 $ and $S_{\pi^0 K_s} > (\sin 2\beta)_{J/\Psi K_s} $ ($S_{\pi^0 K_s} = (\sin 2\beta)_{J/\Psi K_s} $ is realized at $\theta=0$) could be a signal for right-handed currents  \cite{hou}.  
More theoretical studies are needed
in order to determine if this is indeed the case.
In particular, a more thorough analysis of uncertainties due to $O(1/m)$ effects
needs to be undertaken.  For example, power corrections to the dipole operator matrix elements remain to be included.  Furthermore,
the impact on $S_{\phi K_s } ,\,S_{\pi^0 K_s}$ of New Physics in all  of the `left-handed' four-quark operators needs to be thoroughly studied.

\section{Polarization and $C\!P$ violation in $B \to VV$ decays}

\begin{figure}[htb]
\centerline{
\hbox{\hspace{1.2cm}
\includegraphics[width=9truecm]{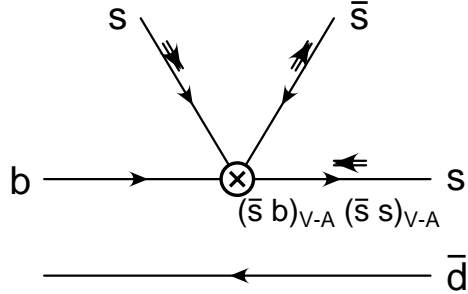}}}
{\caption[1]{\label{fig:factorizable}
Quark helicities (short arrows) for the $\bar B \to \phi K^* $ matrix element of the operator $(\bar s b)_{V-A}  (\bar s s )_{V- A}$ in naive factorization.
Upward lines form the $\phi$ meson.}}
\end{figure}

A discussion of polarization in $B \to VV$ decays has been presented
in \cite{kaganVV} in the framework of QCD factorization.  Here we summarize some of the results.
To begin with we note that the polarization should be sensitive to the $V-A$ 
structure of the Standard Model, due to the power suppression associated with the `helicity-flip' of a collinear quark.  
For example, in the Standard Model the factorizable graphs for
$\bar B\to \phi K^*$ are 
due to transition operators with 
chirality structures $(\bar s b)_{V-A}  (\bar s s )_{V\mp A}$, see Figure \ref{fig:factorizable} .
In the helicity amplitude ${\cal \bar A}^- $ a collinear $s$ or $\bar s$ quark with positive helicity 
ends up in the negatively
polarized $\phi$, whereas in 
${\cal \bar A}^+$ a second quark `helicity-flip{'} is required in the form factor transition.  Collinear quark helicity flips require transverse momentum, $k_\perp$, implying a suppression of $O(\Lambda_{\rm QCD} / m_b  )$ per flip.
In the case of new right-handed currents, e.g.,  $(\bar s b)_{V+A}  (\bar s s)_{V\pm A}$, the helicity amplitude hierarchy would be inverted, with ${\cal \bar A}^+$ and ${\cal \bar A}^-$ requiring one and two helicity-flips,
respectively.

In naive factorization the 
$\bar{B} \to \phi K^{*}$ helicity amplitudes, supplemented by the large energy form factor relations \cite{charlesetal}, satisfy
\begin{equation}
{\cal \bar A}^0 \propto   f_{\phi} m_{B\,}^2  \zeta^{K^*}_\parallel ,~~~
{\cal \bar A}^{-}\propto- f_{\phi} m_{\phi} m_{B\,} 2\, \zeta^{K^*}_\perp\, , 
~~~ {\cal \bar A}^{+}\propto - f_{\phi} m_{\phi} m_{B\,} 2\,  \zeta^{K^*}_\perp r^{K^*}_\perp \,.
\label{Av1v2SCET}
\end{equation}
$\zeta^V_{\parallel }$ and $\zeta^V_{\perp}$ are the $B \rightarrow V$ form factors 
in the large energy limit \cite{charlesetal}.  Both scale as $m_b^{-3/2}$ in the heavy quark limit,
implying 
${\cal\bar  A}^-  /{\cal\bar  A}^0 = O( m_\phi / m_B )$.
$r_\perp$ parametrizes form factor helicity suppression. 
It is given by
\begin{equation}
\label{rperp}
 r_\perp = \frac{(1+m_{V_1}/{m_B} ) 
{ A^{V_1}_1}  - (1- {m_{V_1}}/{m_B} ) { V^{V_1}}     }
{(1+{m_{V_1}}/{m_B} ) 
{ A^{V_1}_1}  + (1- {m_{V_1}}/{m_B} ) { V^{V_1}}    }\,,
\end{equation}
where $A_{1,2} $ and $V$ are the axial-vector and vector 
current form factors, respectively. 
The large energy relations imply that $ r_\perp $ vanishes at leading power,
reflecting the fact that helicity suppression is $ O(1/m_b )$.
Thus, ${\cal\bar  A}^+  /{\cal\bar  A}^- = O(\Lambda_{\rm QCD}  / m_b )$.
Light-cone QCD sum rules  \cite{ballbraun}, and lattice form factor determinations scaled to low $q^2$ using 
the sum rule approach \cite{lattice1}, give 
$r^{K^*}_\perp \approx 1-3 \,\%$; QCD sum rules give $r^{K^*}_\perp \approx 5\,\%$ \cite{fazio}; and the BSW model gives $r^{K^*}_\perp \approx 10\%$ \cite{BSW}.

The polarization fractions in the transversity basis (\ref{transversity})
therefore satisfy
\begin{equation}
 \label{SMpred}
1-f_L = {\cal  O}\left({1 / m_b^2}\right),
~~~
{f_\perp/ f_\parallel  }=  1+ {\cal  O} \left({1/ m_b}\right),
\end{equation} 
in naive factorization, where the subscript $L$ refers to longitudinal polarization,  $f_i = \Gamma_i /\Gamma_{\rm total}$, and $f_L + f_\perp+f_\parallel =1$.
The measured longitudinal fractions for $B \to \rho \rho$ 
are close to 1 \cite{BelleRhopRho0,BaBarRhopRho0}.
This is not the case for $B \rightarrow \phi K^{*0}$ for which full angular analyses
yield 
\begin{eqnarray}
\label{fLperpphiKst}
f_L &=& .43 \pm .09 \pm .04,~~~f_\perp = .41 \pm.10\pm .04~~\mbox{\cite{BellePhiKst}}\\
f_L &=& .52 \pm .07 \pm .02,~~~f_\perp = .27 \pm.07 \pm .02~~\mbox{\cite{BaBarPhiKst}}.
\end{eqnarray}
Naively averaging the Belle and BaBar measurements (without taking correlations into account) yields 
$f_\perp /f_\parallel = 1.39 \pm .69$. 
We must go beyond naive factorization in order to determine if the small value of $f_L (\phi K^* )$ could simply be due to the dominance of QCD penguin operators in $\Delta S=1$ decays, rather than New Physics.  In particular, it is necessary to determine if the power counting in (\ref{SMpred}) is preserved by non-factorizable graphs, i.e., penguin contractions, vertex corrections, spectator interactions, annihilation graphs, and graphs involving higher Fock-state gluons.  This question can be addressed in 
QCD factorization \cite{kaganVV}.

In QCD factorization exclusive two-body decay amplitudes are given in terms 
of convolutions of hard scattering kernels with meson light-cone distribution amplitudes \cite{BBNS}--\cite{benekeneubert}.
At leading power this leads to factorization of short and long-distance physics.
This separation breaks down at sub-leading powers
with the appearance of logarithmic infrared divergences, e.g.,
$\int_{0}^1 d x /x \sim \ln m_B /\Lambda_h $, where $x$ is the light-cone quark momentum fraction in a final state meson, and $\Lambda_h  \sim \Lambda_{\rm QCD}$ is a physical infrared cutoff.
Nevertheless, the power-counting for all amplitudes can be obtained.  The extent to which it holds numerically can 
be determined by assigning large uncertainties to the logarithmic divergences. 
Fortunately, certain polarization observables are less sensitive to this uncertainty, particularly after 
experimental constraints, e.g., total rate or total transverse rate, are imposed.

\begin{figure}[htb]
\centerline{
\hbox{$\begin{array}{cc}
\includegraphics[width=7.5cm]{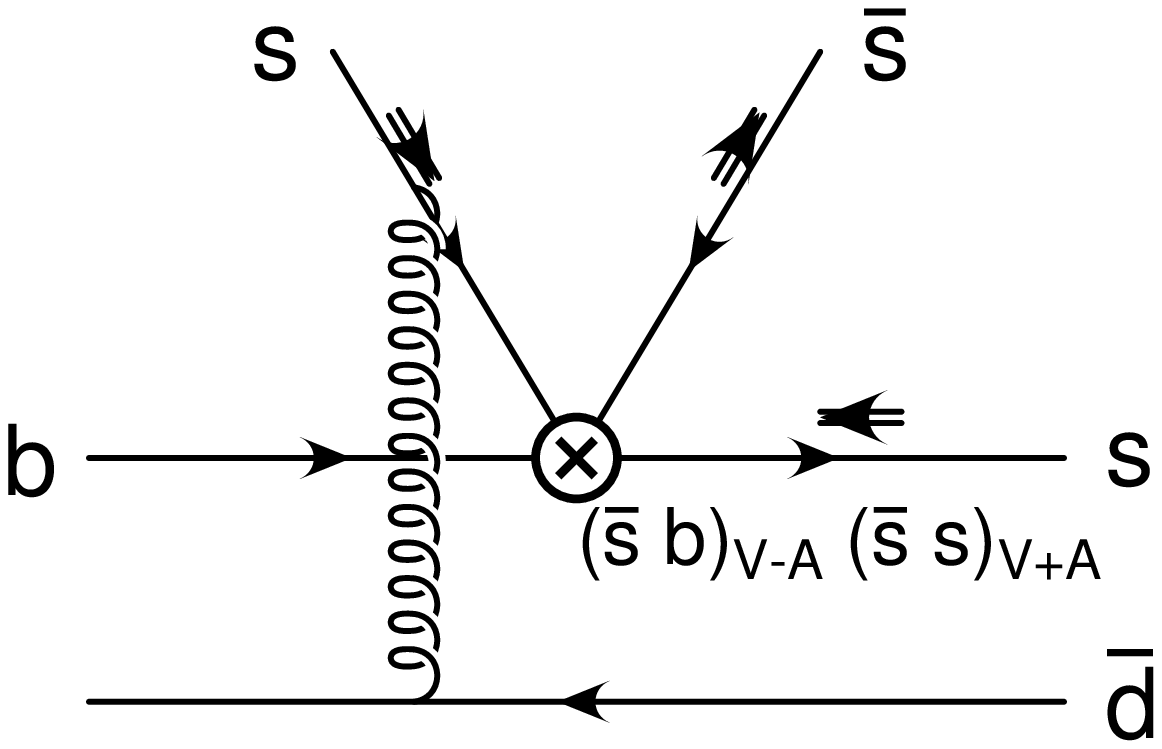} &
\hspace{-.3cm} \includegraphics[width=7.5cm]{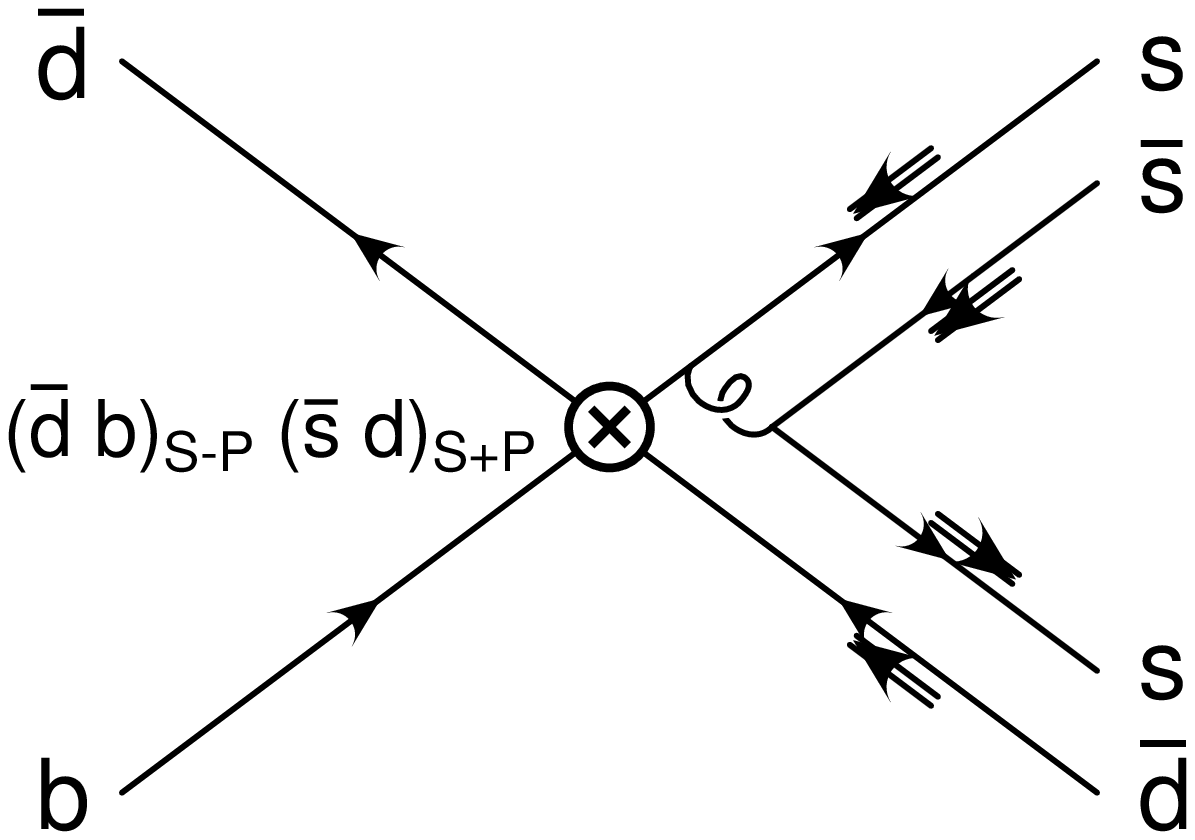}
\end{array}$}}
{\caption[1]{\label{fig:annihilation}
Quark helicities in $\bar B \to \phi K^*$ matrix elements: the hard spectator interaction for the operator $(\bar s b)_{V-A}  (\bar s s )_{V\mp A}$ (left), and  annihilation graphs for the operator $(\bar d b)_{S-P}  (\bar s d )_{S+P}$ with gluon emitted from the final state quarks (right).}}
\end{figure}

Examples of logarithmically divergent hard spectator interaction and QCD penguin annihilation graphs are shown in Figure \ref{fig:annihilation}, with
the quark helicities indicated.
The power counting for the helicity amplitudes of the annihilation graph, including logarithmic divergences,
is
\begin{equation}
\label{A3fpowers}
{\cal \bar A}^0 ,~{\cal \bar A}^- =  { O\left( {1\over m^2}{ \,{\ln}^2\!{m \over \Lambda_h }}\right)},\qquad 
{\cal \bar A}^+ = { O\left( {1\over m^4}{ \,{\ln}^2\!{m \over \Lambda_h }}\right)}\,.
\end{equation}
The logarithmic divergences are associated with the limit in which both the $s$ and $\bar s$ quarks originating from the gluon are soft.
The annihilation topology implies an overall factor of $1/m_b$.
Each remaining factor of $1/m_b $ is associated with a quark helicity flip.
In fact, adding up all of the helicity amplitude contributions in QCD factorization {\it formally} preserves the naive factorization power counting in (\ref{SMpred}) \cite{hawaiitalk,kaganVV}.  
Recently, the first relation in (\ref{SMpred}) has been confirmed 
in the soft collinear effective theory \cite{ianetal}.
However, as we will see below, it need not hold numerically 
because of QCD penguin annihilation.


\subsection{Numerical results for polarization}

The numerical inputs are given in \cite{kaganVV}.  The logaritmic divergences are modeled as in \cite{BBNS3,benekeneubert}.  For example, in the annihilation amplitudes the quantities $X_A$ are introduced 
as
\begin{equation}\label{XA} \int_0^1 \frac{dx}{x} \rightarrow  X_{A}  = (1 + {\varrho_A } e^{i{ \varphi_A}} )   \ln\frac{m_B}{\Lambda_h}\, ;~~~
  { \varrho_A} \le 1 \,,~~~ \Lambda_h \approx 0.5\, {\rm GeV}\,.
  \end{equation}
This parametrization reflects the physical $O(\Lambda_{\rm QCD} )$
cutoff, and allows for large strong phases $\varphi_A \in [0,2 \pi]$ from soft rescattering.
The quantities $X_A$ (and the corresponding hard spectator interaction 
quantities $X_H$) are varied independently for unrelated convolution integrals.

The predicted longitudinal polarization fractions $f_L (\rho^- \rho^0 )$ and 
$f_L (\rho^- \rho^+ )$ are close to unity, in agreement 
with observation  \cite{BelleRhopRho0,BaBarRhopRho0}
and with naive power counting (\ref{SMpred}).
The theoretical uncertainties are small, particularly after imposing the 
branching ratio constraints, due to the absence of (for $\rho^- \rho^0$) or CKM suppression of (for $\rho^- \rho^+ $) the QCD penguin amplitudes.

\begin{figure}[htb]
\centerline{
\hbox{\hspace{-.4cm}
\includegraphics[width=6.truecm]{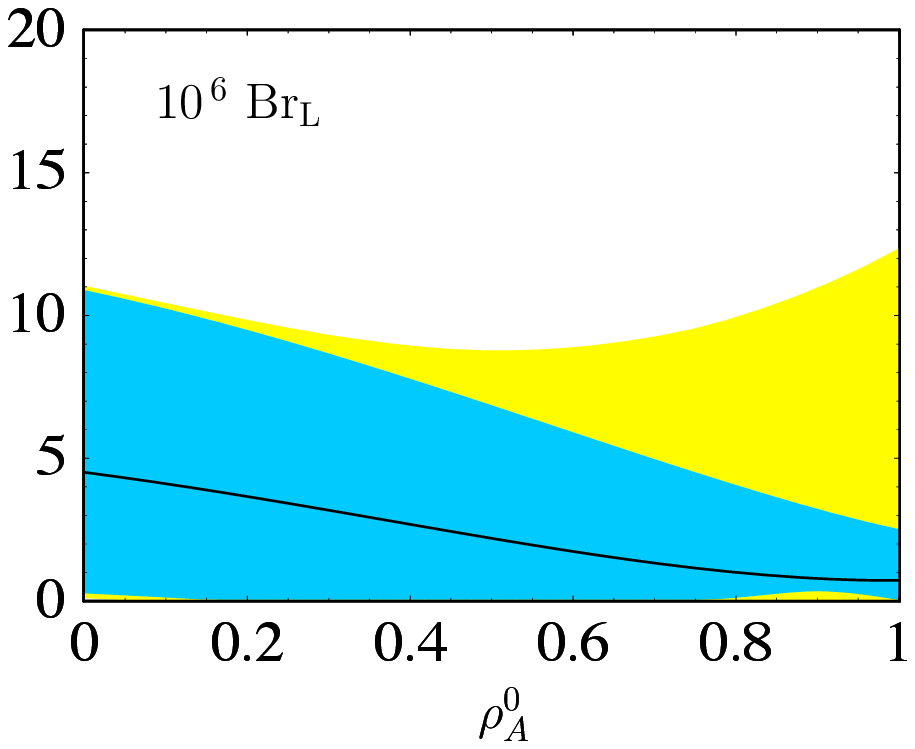}\hspace{1.0cm} 
\includegraphics[width=6.truecm]{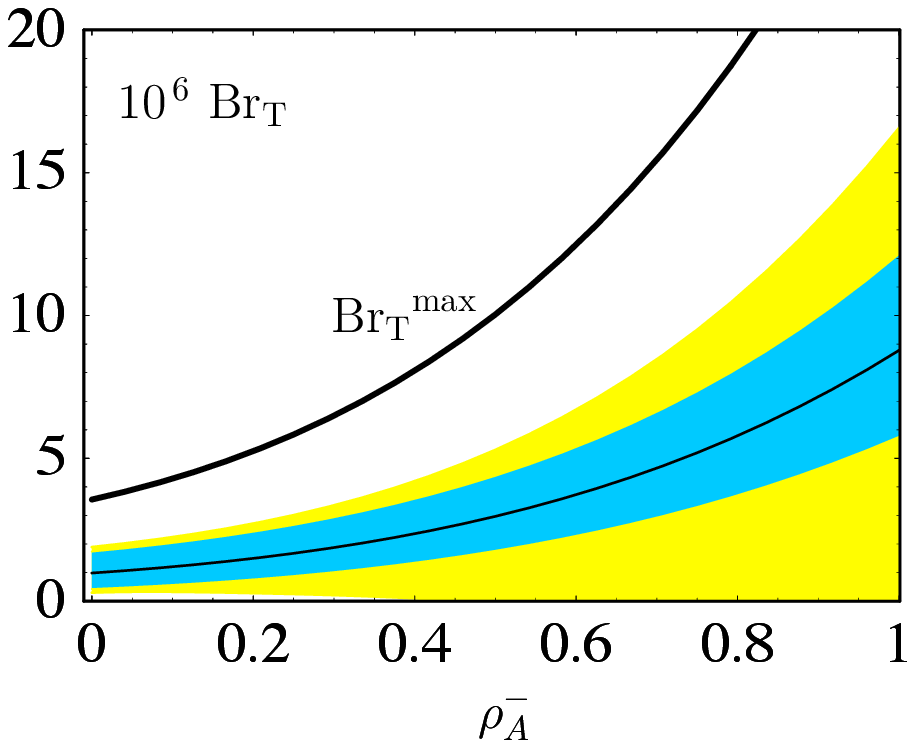}}}
{\caption[1]{\label{fig:phiKstpolarization}
${\rm Br}_{L} ( \phi K^{*0})$ vs.  $\rho^{0}_A $ (left), ${\rm Br}_{T} ( \phi K^{*0})$ vs.  $\rho^{-}_A $ (right).
Black lines: default inputs. Blue bands:
input parameter variation uncertainties added in quadrature, keeping 
default annihilation and hard spectator interaction parameters.  Yellow bands: additional uncertainties, added in quadrature, from variation of
parameters entering logarithmically divergent annihilation and hard spectator interaction power corrections.  Thick line: ${\rm Br}^{\rm max}_T$ under simultaneous variation of all inputs.}}
\end{figure}

Averaging the Belle and BaBar $\bar B\to \phi K^{*0}$ measurements \cite{BellePhiKst,BaBarPhiKst,BaBarRhopRho0} yields 
$f^{\rm exp}_L=.49\pm .06$ and ${\rm Br }^{\rm exp}=10.61\pm1.21$, or 
 ${\rm Br}^{\rm exp}_L = 5.18 \pm .86$ and ${\rm Br}^{\rm exp}_T = 5.43 \pm .88$.
${\rm Br}_L$ and ${\rm Br}_T ={\rm Br}_\perp +  {\rm Br}_\parallel $ are the $C\!P$-averaged 
longitudinal and total transverse branching ratios, respectively.
In the absence of annihilation, the predicted branching ratios are
$10^6\, {\rm Br}_L = 5.15^{+6.79+.88}_{-4.66-.81} $
and $10^6 \,{\rm Br}_T = .61^{+.60+.38}_{-.42-.29} $,
where the second (first) set of error bars is due to variations of $X_H$ (all other inputs).
However, the $(S+P ) (S-P)$ QCD penguin annihilation graph in  Figure \ref{fig:annihilation} can play an important role in both ${\cal \bar A}^0$ and ${\cal \bar A}^-$ due to the appearance of a logarithmic divergence squared ($X_A^2 $), the large Wilson coefficient $C_6$, and a $1/N_c $ rather than $1/N_c^2 $ dependence.
Although formally $O(1/m^2 )$, see (\ref{A3fpowers}), these contributions can be $O(1)$ numerically.   This is illustrated in Figure \ref{fig:phiKstpolarization},
where ${\rm Br}_L $ and ${\rm Br}_T$ are plotted versus the quantities $\rho_A^0 $ and $\rho_A^- $, respectively, for 
$\bar B \to \phi K^{*0}$.  
$\rho_A^0 $ and $\rho_A^- $ enter the parametrizations (\ref{XA})
of the logarithmic divergences appearing in the longitudinal and negative helicity $(S+P ) (S-P)$ annihilation amplitudes, respectively.
As $\rho_A^{0,-}$ increase from 0 to 1, the 
corresponding annihilation amplitudes increase by more than an order of magnitude.  The theoretical uncertainties on the rates are very large.
Furthermore, the largest input parameter uncertainties in 
${\rm Br}_L$ and ${\rm Br}_T$ are a priori unrelated.
Thus,  it is clear from Figure \ref{fig:phiKstpolarization} that the QCD penguin annihilation amplitudes can account for the $\phi K^{*0}$ measurements.  
Similarly, the BaBar measurement of $f_L (\phi K^{*-} ) \approx 50\%$ \cite{BaBarRhopRho0} can be accounted for.

Do the QCD penguin annihilation amplitudes also imply large transverse
polarizations in $B \to \rho K^* $ decays?  The answer depends on 
the pattern of $SU(3)_F$ flavor symmetry violation in these amplitudes.
For light mesons containing a single strange quark, e.g., $K^*$,
non-asymptotic effects shift the weighting of the meson distribution amplitudes towards larger strange quark momenta.  
As a result, the suppression of $s \bar s $ popping relative to light quark popping in annihilation amplitudes can be $O(1)$, which is consistent 
with the order of magnitude hierarchy
between the $\bar B \to D^0 \pi^0 $ and $\bar B \to D_s^+ K^- $ rates ~\cite{pirjolstewart}.
(See \cite{benekesu3} for a discussion of other sources of $SU(3)$ violation).
In the present case, this implies that the longitudinal polarizations should satisfy $f_L ( \rho^\pm K^{*0}) \lsim f_L (\phi K^{*})$ in the Standard Model \cite{kaganVV}.
Consequently, $f_L ( \rho^\pm K^{*0}) \approx 1$ would suggest 
that $U$-spin violating New Physics entering mainly in the $b\to s \bar s s$ channel 
is responsible for the small $f_L (\phi K^{*} )$.
One possibility would be right-handed vector currents; they could interfere constructively (destructively) in ${\cal\bar A}_\perp$ (${\cal \bar A}_0$) transversity amplitudes, see (\ref{ANPVV}).
Alternatively, a parity-symmetric scenario would only affect ${\cal\bar A}_\perp$.
A more exotic possibility would be tensor currents; they would contribute to the longitudinal and transverse amplitudes at sub-leading and 
leading power, respectively, opposite to the vector currents.

We should mention that our treatment of the charm (and up) quark loops in the penguin amplitudes follows the usual perturbative approach used in QCD factorization \cite{BBNS}--\cite{benekeneubert}.  The authors of \cite{ianetal} believe that the region of phase space in which the 
charm quark pair has invariant mass $q^2 \sim 4 m_c^2$, and is thus moving non-relativistically, should be separated out into a long-distance
`charming penguin' amplitude \cite{ciuchinicharm}.   NRQCD arguments are invoked to claim that 
such contributions are $O(v) $, where $v \approx .4-.5$, so that they could effectively be of leading power.  Furthermore, it is claimed that  the transverse components may also be of leading power, thus potentially accounting for  $f_L (\phi K^* )$.  However, a physical mechanism by which a collinear quark helicity-flip could arise in this case without power suppression remains to be clarified.
Arguments against a special treatment of this region of phase space \cite{BBNS,BBNS3} are based on parton-hadron duality.  It should be noted that in QCD factorization this region of $q^2$ contributes negligibly to the $B \to VV$ penguin amplitudes, particularly in the transverse components.
More recently, the low value of $f_L (\phi K^* )$ has been addressed using 
a purely hadronic model for soft rescattering of intermediate two-body charm states, i.e., 
$B \to D^{(*)}_{s} D^{(*)} \to \phi K^* $ \cite{defazio}.
This approach has been criticized previously on the grounds that a "purely hadronic language, suitable for kaon decays" is not applicable to the case of $B$ decays, where the "number of channels, and the energy release are large" \cite{BBNS3}.  In particular,
many intermediate multi-body channels have been ignored which are predicted to lead to systematic
amplitude cancelations in the heavy quark limit.

\subsection{A test for right-handed currents}

\begin{figure}[htb]
\centerline{
\hbox{\hspace{-.4cm}
\includegraphics[width=6.truecm]{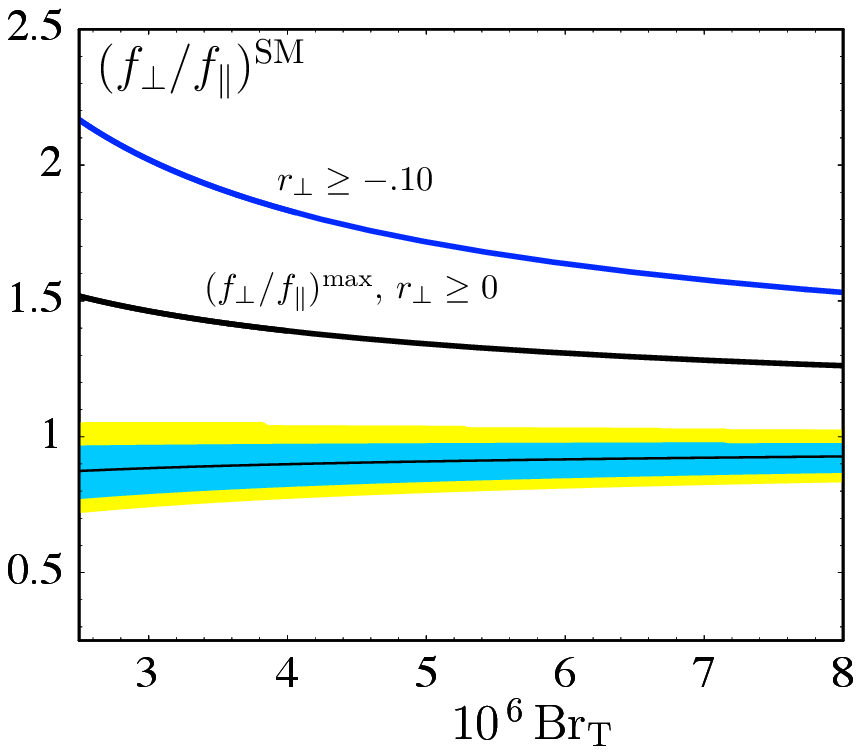}\hspace{.5cm} 
\includegraphics[width=6.truecm]{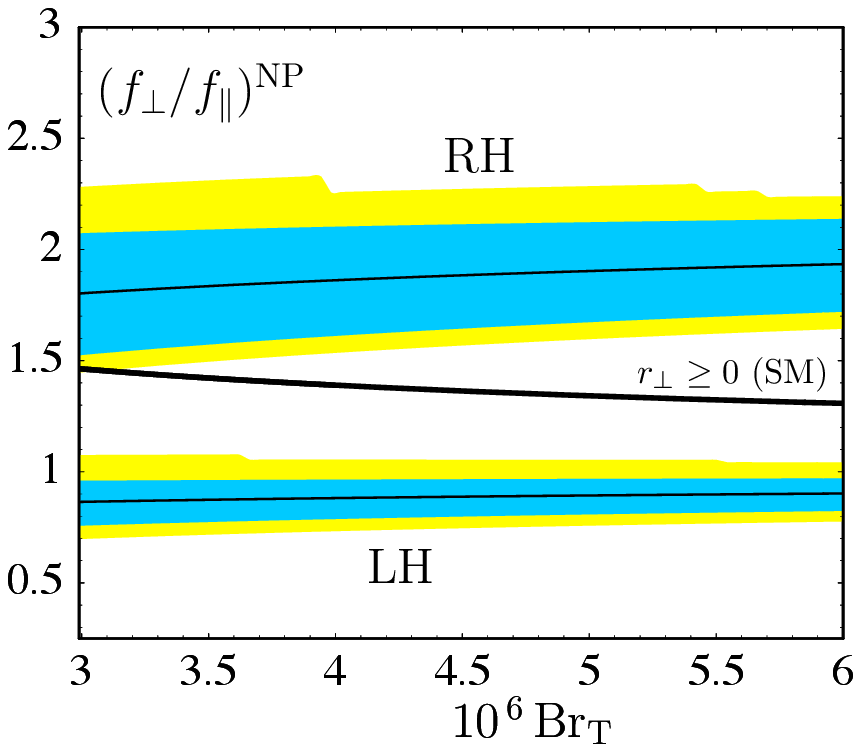}}}
{\caption[1]{\label{fig:fperppar}
$f_\perp /f_\parallel $ vs. ${\rm Br}_T$ in the SM (left), and with new
RH or LH currents (right).  Black lines, blue bands, and yellow bands are as in Figure \ref{fig:phiKstpolarization}.  Thick lines: $(f_\perp /f_\parallel)^{\rm max}$ in the Standard Model for indicated ranges of $r^{K^*}_\perp$ under simultaneous variation of all inputs. Plot for $r^{K^*}_\perp >0$ corresponds to ${\rm Br}_T^{\rm max}$ in Figure \ref{fig:phiKstpolarization}.}}
\end{figure}

Does the naive factorization relation 
$f_\perp /f_\parallel = 1 + O(\Lambda_{\rm QCD} /m_b )$ (\ref{SMpred}) survive
in QCD factorization?  This ratio is very sensitive to the quantity 
$r_\perp $ defined in  (\ref{rperp}).  
As $r_\perp$ increases, $f_\perp /f_\parallel$ decreases.
The range $r^{K^* }_\perp = .05 \pm .05$ spanning existing model determinations \cite{ballbraun}--\cite{BSW} is taken in \cite{kaganVV}. 
In Figure \ref{fig:fperppar} (left) the resulting predictions for $f_\perp /f_\parallel $ and ${\rm Br}_T$ 
are studied simultaneaously for $\bar B \to \phi K^{*0}$ in the Standard Model.
Note that the theoretical uncertainty for $f_\perp /f_\parallel $ is much smaller 
than for $f_L$.
Evidently, the above relation still holds, particularly at larger values of ${\rm Br}_T$ where QCD penguin annihilation dominates both ${\rm Br}_\perp $ and ${\rm Br}_\parallel$.

A ratio for $f_\perp /f_\parallel $ in excess of the Standard Model range, e.g.,  
$f_\perp /f_\parallel > 1.5$ if $r_\perp > 0$,
would signal the presence of new right-handed currents.  
This is due to the inverted hierarchy between ${\cal \bar A}^-$ and ${\cal \bar A}^+$  for right-handed currents, and is reflected in the sign difference with which the Wilson coefficients $\tilde{C}_i$ enter 
${\cal \bar A}_\perp $ and ${\cal \bar A}_\parallel$.
For illustration, new contributions to the QCD penguin operators are considered in Figure \ref{fig:fperppar} (right).
At the New Physics matching scale $M$, these can be parametrized as 
${\stackrel{\scriptscriptstyle{(\sim)}}{C_4}} = {\stackrel{\scriptscriptstyle{(\sim)}}{C_6}} = - 3\, {\stackrel{\scriptscriptstyle{(\sim)}}{C_5}} = -3\, {\stackrel{\scriptscriptstyle{(\sim)}}{C_3}}={\stackrel{\scriptscriptstyle{(\sim)}}{\kappa}} $.
For simplicity, we take $M\!\approx\! M_W$ and consider two cases:
$\kappa = -.007 $ or new left-handed currents (lower bands), and $\tilde{\kappa}=-.007$ or new right-handed currents (upper bands),
corresponding to $C^{NP}_{4\, (5)} (m_b)$ or $\tilde{C}^{NP}_{4 \,(5)} (m_b) \approx  .18\,  C_{4\, (5)}^{SM}  (m_b)$, and
$C^{NP}_{6 \,(3)} (m_b)$ or
$\tilde{C}^{NP}_{6\, (3)} (m_b) \approx  .25  \,C_{6 \,(3)}^{SM}  (m_b)$.
Clearly,
moderately sized right-handed currents could increase $f_\perp /f_\parallel$ well beyond the Standard Model
range if $r_\perp \ge 0$.
However, new left-handed currents would have little effect.

\subsection{Distinguishing four-quark and dipole operator effects}

\begin{figure}[t]
\centerline{
\hbox{\hspace{0.2cm}
\includegraphics[width=4.5truecm,height=2.8truecm]{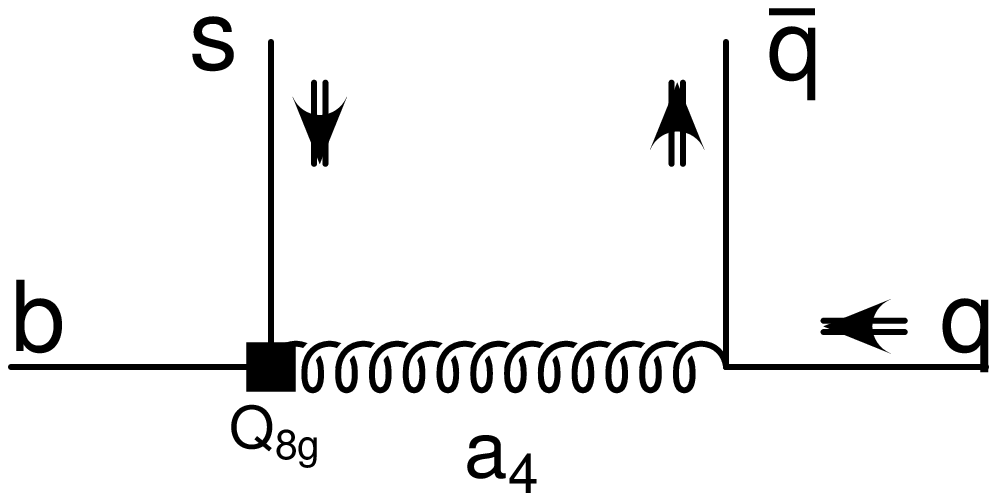}\hspace{1.5cm} 
\includegraphics[width=4.5truecm,height=2.8truecm]{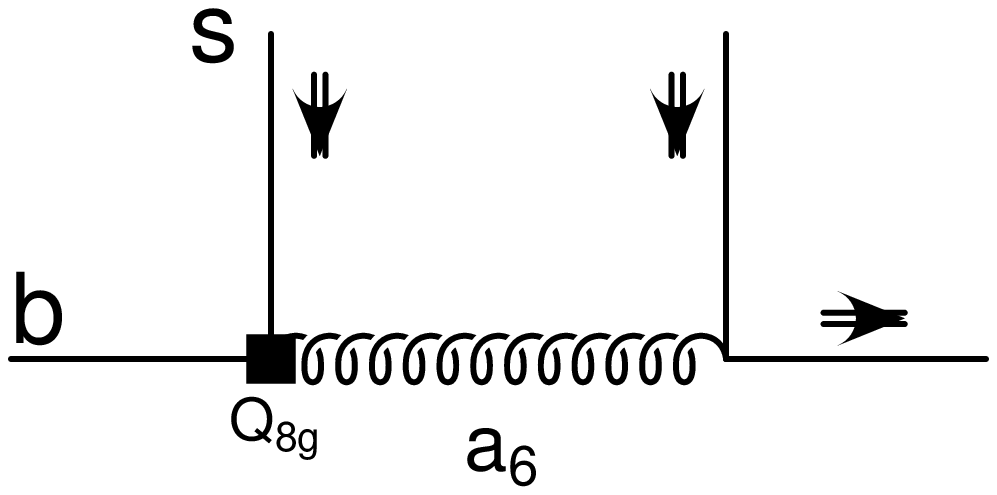}}}
{\caption[1]{\label{fig:Q8g}
Quark helicities for the $O(\alpha_s)$ penguin contractions of $Q_{8g}$.
The upward lines form the $\phi$ meson in $\bar B \to \phi K^*$ decays.}}
\end{figure}

The $O(\alpha_s )$ penguin contractions of the chromomagnetic dipole operator $Q_{8g}$ are illustrated in Figure \ref{fig:Q8g}.   
$a_4$ and $a_6$ are the QCD factorization coefficients of the transition operators $(\bar q b)_{V-A} \otimes (\bar D q)_{V-A} $ and $ (\bar q b)_{S-P} \otimes (\bar D q)_{S+P} $, respectively, where $q$ is 
summed over $u,d,s$ \cite{BBNS3,benekeneubert}.
Only the contribution on the left ($a_4$) to the longitudinal helicity amplitude ${\cal \bar A}^0$ is non-vanishing \cite{kaganVV}.  In particular, the chromo- and electromagnetic dipole operators $Q_{8g}$ and $Q_{7\gamma}$ {\it do not contribute to the 
transverse penguin amplitudes} at $O(\alpha_s )$ due to angular momentum conservation:  the dipole tensor current couples to a transverse gluon, but a `helicity-flip' for $q$ or $\bar q$ 
in Figure \ref{fig:Q8g} would require a longitudinal gluon coupling.
Formally, this result follows from Wandura-Wilczek type relations 
among the vector meson distribution amplitudes,
and the large energy relations between the tensor-current and vector-current form factors.  
Transverse amplitudes in which a vector meson contains a collinear higher Fock state gluon also vanish at ${\cal O}(\alpha_s)$, as
can be seen from 
the vanishing of the corresponding partonic dipole operator graphs in the same momentum configurations.
Furthermore, the transverse ${\cal O}(\alpha_s^2 )$ contributions involving spectator interactions are highly suppressed.

This 
has important implications for New Physics searches.  
For example, in pure penguin decays to CP-conjugate final states $f$, e.g.,
$\bar B\to \phi\, ( K^{*0} \to  K_s \pi^0 )$, if the transversity basis time-dependent CP asymmetry parameters $(S_f)_{\perp} $ and $(S_f)_{\parallel} $ are consistent with  $(\sin 2 \beta)_{J/\psi K_s}$, and $(S_{f})_0$ is not,
then this would signal new CP violating contributions to the
chromomagnetic dipole operators.  However, deviations in $(S_{f})_\perp $ or
$(S_{f})_\parallel $ would signal new CP violating four-quark operator contributions.  
If the
triple-products $A_T^0 $ and  $A_T^{\parallel} $ (\ref{AT})
do not vanish and vanish, respectively, 
in pure-penguin decays, then this
would also signal new CP violating contributions to the
chromomagnetic dipole operators.  This assumes that a significant strong phase difference is measured between ${\cal \bar A}_\parallel $ and ${\cal \bar A}_\perp$, for which there is some experimental indication \cite{BaBarPhiKst}.
However, non-vanishing $A_T^\parallel $, or non-vanishing transverse direct CP asymmetries
would signal the intervention of four-quark operators.
The above would help to discriminate between different explanations for 
an anomalous $S_{\phi K_s}$, which fall broadly into two categories:  
radiatively generated dipole operators, 
e.g., supersymmetric loops; or tree-level four-quark operators, e.g., 
flavor changing (leptophobic) $Z^\prime $ exchange \cite{berger}, $R$-parity violating couplings \cite{Datta}, or color-octet exchange \cite{burdman}.  Finally, a large value for $f_\perp /f_\parallel $ would be a signal for right-handed {\it four-quark} operators.

\section{Conclusion}

There are a large number of penguin-dominated  
rare hadronic $B$ decay modes in the Standard Model in which departures from 
null $C\!P$ asymmetry predictions would be a signal for New Physics.
We have seen that in order to detect the possible intervention of 
new $b \to s_R$ right-handed currents it is useful to organize these modes according to the parity of the final state. 
$SU(2)_L \times SU(2)_R \times U(1)_{B-L} \times P$ symmetric models in which new $C\!P$ violating contributions to the effective $\Delta B=1$ Hamiltonian
are to good approximation parity symmetric at the weak scale, would only give rise to significant deviations from null $C\!P$ asymmetries in parity-odd final states.  For example, no deviations from the null Standard Model $C\!P$ asymmetry predictions in 
$S_{\eta^\prime K_s}$, $(S_{\phi K^{*0}})_{0, \parallel} $, $A_{CP} (\phi K^{*\pm})_{0,\parallel}$, $A_{CP} (K^0 \pi^\pm)$ could be accompanied by significant
deviations in $S_{\phi K_s}$, $A_{CP} (\phi K^\pm)$, $A_{CP} (\phi K^{*\pm})_{\perp}$
$(S_{\phi K^{*0}})_{\perp} $,  and $S_{f^0 K_s} $.  
This would provide a clean signal for left-right symmetry.
However, the precision of $C\!P$ asymmetry measurements necessary 
to discern the existence of such a pattern would require 
a high luminosity $B$ factory.
Remarkably, approximate parity invariance in the $\Delta B=1 $
effective Hamiltonian can be realized even if the $SU(2)_R$ symmetry 
breaking scale $M_R$ is as large as $M_{\rm GUT}$.  An explicit example in which large departures from the null predictions are possible,
but in which deviations from parity invariance can be as small as $O(1\%)$
for $M_R \le M_{\rm GUT}$, is provided 
by squark-gluino loops in parity-symmetric SUSY models.  
It is noteworthy that, due to parity invariance, stringent bounds 
on new sources of  $C\!P$ and flavor violation arising from the 
$^{199}$Hg mercury edm are naturally evaded in such models.

More generally, in models in which new contributions to 
Standard Model (left-handed) and opposite chirality (right-handed) effective operators are
unrelated,  the $C\!P$ asymmetries in the $P$-odd and $P$-even null Standard Model modes could differ substantially both from each other, and from the null predictions.  This is because the right-handed operator Wilson coefficients enter with opposite sign in the amplitudes for decays to $P$-odd and $P$-even final states.
Unfortunately, $C\!P$ asymmetry predictions 
have large theoretical uncertainties due to $1/m$ power corrections, especially 
from the QCD penguin annihilation amplitudes.  We therefore can not
rule out substantial differences between new  $C\!P$ violating effects in parity-even and parity-odd modes arising solely from left-handed currents.
However, very large differences, e.g., $S_{\phi K_s } < 0 $ and $S_{\pi^0 K_s} > (\sin 2\beta)_{J/\Psi K_s} $, may provide a signal for $C\!P$ violating right-handed
currents.  More theoretical work will be required in order to make this statement more precise.

Polarization measurements in $B$ decays to light vector meson pairs offer a unique opportunity to probe the chirality structure of rare hadronic $B$ decays.
A Standard Model analysis which includes all non-factorizable graphs in QCD factorization shows that the longitudinal polarization formally satisfies $1-f_L = {\cal O}(1/m^2)$, as in naive factorization.  However, the contributions of a particular QCD penguin annihilation graph
which is formally ${\cal O}(1/m^2 )$ can be ${\cal O}(1)$ numerically in longitudinal and 
negative helicity $\Delta S\!=\!1$ $\bar B$ decays.  Consequently, the observation of $f_L (\phi K^{*0,-} ) \approx 50\%$ can be accounted for, albeit with large theoretical errors. 
The expected pattern of $SU(3)_F$ violation in the QCD penguin annihilation graphs, i.e., large suppression of $s\bar s$ relative to 
$u \bar u $ or $d\bar d$ popping, implies that the longitudinal polarizations should satisfy $f_L ( \rho^\pm K^{*0}) \lsim f_L (\phi K^{*})$ in the Standard Model.
Consequently, $f_L ( \rho^\pm K^{*0}) \approx 1$ would suggest
that $U$-spin violating New Physics entering mainly in the $b\to s \bar s s$ channel is responsible for the small values of $f_L (\phi K^{*} )$.

The ratio of transverse rates in the transversity basis satisfies
$\Gamma_\perp /\Gamma_\parallel = 1 + {\cal O}(1/m )$, in agreement with naive power counting.
A ratio in excess of the predicted Standard Model range would signal the presence of new right-handed currents in dimension-6 four-quark operators.  
The maximum ratio attainable in the Standard Model is sensitive to the $B \to V$ form factor combination $r_\perp$, see (\ref{rperp}), which controls helicity suppression in form factor transitions.  All existing model determinations give a positive sign for $r_\perp$, which would imply
$\Gamma_\perp (\phi K^{*} ) /\Gamma_\parallel (\phi K^{*} )  < 1.5 $ in the Standard Model.
The magnitude and especially the sign of $r_\perp^{K^*}$
is clearly an important issue which should be clarified further with dedicated lattice studies.

Contributions of the dimension-5 $b \to sg$ dipole operators to the transverse $B \to VV$ modes are highly suppressed, due to angular momentum conservation.
Comparison of $C\!P$ violation 
involving the longitudinal modes with $C\!P$ violation only involving the transverse modes in pure penguin $\Delta S=1$ decays could therefore distinguish between new contributions to the dipole  and four-quark operators.
More broadly, this could distinguish between scenarios in which New Physics effects are loop induced and scenarios in which they are tree-level induced, as it is difficult to obtain ${\cal O}(1)$ CP-violating effects from dimension-6 operators beyond tree-level.  Again, a high luminosity $B$ factory will be required in order to obtain the necessary level of precision in $C\!P$
violation measurements.

\newpage
{\it Acknowledgments:} 
I would like to thank Martin Beneke, Alakhabba Datta, Keith Ellis, Guy Engelhard, Andrei Gritsan, Yuval Grossman, Matthias Neubert , Uli Nierste, Yossi Nir, Dan Pirjol, and Ian Stewart for useful discussions.  This work was supported by the Department of Energy under Grant DE-FG02-84ER40153.

\def\Discussion{
\setlength{\parskip}{0.3cm}\setlength{\parindent}{0.0cm}
     \bigskip\bigskip      {\Large {\bf Discussion}} \bigskip}
\def\speaker#1{{\bf #1:}\ }
\def\endDiscussion{}

\end{document}